# Few-electron molecular states and their transitions in a single InAs quantum dot molecule

T. Ota[1*], M. Rontani[2], S. Tarucha[1,3],
Y. Nakata[4], H. Z. Song[4], T. Miyazawa[4], T. Usuki[4], M. Takatsu[4], and N. Yokoyama[4]

[1]*ERATO/SORST/ICORP, JST, Atsugi-shi, Kanagawa 243-0198, Japan*
[2]*INFM-CNR National Research Center S3, Via Campi 213/A, 41100 Modena, Italy*
[3]*Department of Applied Physics, University of Tokyo, Bunkyo-ku, Tokyo, 113-0033, Japan*
[4]*Fujitsu Laboratories Ltd., Atsugi-shi, Kanagawa 243-0197, Japan*

**Abstract**

We study electronic configurations in a *single pair* of vertically coupled self-assembled InAs quantum dots, holding just a few electrons. By comparing the experimental data of non-linear single-electron transport spectra in a magnetic field with many-body calculations, we identify the spin and orbital configurations to confirm the formation of molecular states by filling both the quantum mechanically coupled symmetric and anti-symmetric states. Filling of the anti-symmetric states is less favored with increasing magnetic field, and this leads to various magnetic field induced transitions in the molecular states.

*Corresponding author.
Present address:
Quantum Solid-State Physics Research Group, NTT Basic Research Laboratories, 3-1, Morinosato Wakamiya, Atsugi-shi, 243-0198, Japan

*E-mail address:* ota@will.brl.ntt.co.jp

Among the various kinds of quantum dot systems, self-assembled quantum dots (SADs) have several advantages such as confinement of electrons and/or holes by heterojunction barriers (depending on bandedge lineups of two semiconductors), no damage by processing, and large level spacing and large interaction energy because of their reduced size. These novel features have stimulated interest for applications to optical and electrical devices as well as fundamental studies on confined interacting electrons and/or holes. Optical spectroscopy techniques have often been employed to characterize the electronic properties of InAs SADs but suffer from inhomogeneous broadening due to the measurement of large SAD ensembles. Certain techniques have also allowed individual SADs to be probed such as optical spectroscopy of restricted regions including just a single SAD [1], scanning tunneling microscopy [2], and transport measurements of single-electron tunneling devices [3]. Atom-like $s$-, $p$-, $d$-orbital states have already been well characterized for single InAs quantum dots (QDs). In particular for specific single-electron tunneling devices, the number $N$ of surplus electrons in the dots can be varied one-by-one, starting from zero. This allows for characterizing the electronic configurations in a controlled manner [4].

Knowledge of atom-like states in single SADs naturally leads to interest in molecular states of coupled QDs, or artificial molecules, because molecular states can be formed by coupling the atom-like states in two dots. The molecular states can be formed by filling both tunnel-coupled symmetric and anti-symmetric states, and the resulting electronic configuration is also characterized by dipolar charge coupling and exchange spin coupling between electrons in the two dots. The couplings are realized in a manner similar to that of real diatomic molecules but with much more freedom than in the "natural" case. Actually, artificial molecule states can be controlled by appropriately designing parameters of the dot system or as a function of gate voltage and magnetic field. In the scheme of self-assembled growth, two dots can align themselves in the vertical direction with abrupt heterojunction barriers [5]. Note that, lithographically defined QDs, the orbital states are weakly electrostatically confined, therefore the coupling between two dots can be significantly modified when tuning either the number of electrons or the inter-dot coupling strength.

Recently, molecular states were studied for a single InAs SAD molecule by optical spectroscopy [6]. The measured spectra arise from electron-hole pairs, and are useful for deriving the single-particle energy spacing between the molecular orbital states. However, there have been no experiments performed to determine the filling of quantum mechanically coupled symmetric (S) and anti-symmetric (AS) electronic states. In this Letter, we use single-electron tunneling spectroscopy to study the electronic configurations in a single InAs QD molecule, holding just a few electrons. We probe just a *single* QD molecule among an ensemble of QD molecules, embedded in a sub-micron size mesa, by depleting all but one of the QD molecules with a side gate voltage [4]. The number $N$ of electrons in the last single QD molecule in the conducting channel is precisely tuned one-by-one, starting from $N=0$. We measure the energy spectra of the ground and excited states as a function of magnetic field ($B$ field) for strongly and weakly coupled QD molecules. We find that the measured spectra compare well with many-body numerical calculations based on standard Fock-Darwin (F-D) states. Note that, in previous experiments [7], the energy spectra of single InAs dots were well reproduced by a F-D state-based calculation. We can also distinguish the filling of the molecular ground states (GSs). We can also realize novel transitions in the molecular GSs, induced by $B$ field.

For the present study we used two materials, A and B, both of which include vertically coupled InAs QDs, grown by molecular beam epitaxy. Material A (B) has a distance between the two InAs wetting layers of 6.5 nm (11.5 nm). The tunnel coupling between the dots is much stronger for material A than for material B. The height and diameter of the dots in both materials are 2-3 nm and 20-30 nm, respectively. The average density of the dots is of the order of $10^9$ cm$^{-2}$. We fabricated sub-micron size single-electron transistors, i.e. devices A (B), using material A (B). These devices consist of a circular mesa whose geometrical diameter is 0.35 µm, surrounded by a Schottky gate metal [8]. Just a few QD molecules exist in the mesa. We measure a current, $I$, flowing vertically through the mesa as a function of gate voltage, $V_G$, and source-drain voltage, $V_{SD}$. The devices are placed in a dilution refrigerator with a bath temperature below 100 mK.

Although a few QD molecules exist in the mesa, we are usually able to observe only a *single* QD molecule near pinch-off point. It is because the number of the QD molecules in the conducting channel reduces due to that the depletion region squeezes the conducting channel in the mesa when the gate voltage is made more negative. Then, we are able to obtain clean results on the last single QD molecule in the conducting channel in the mesa. We initially measure the non-linear transport spectra, i.e. *dI/dV$_{SD}$* in the *V$_{SD}$* and *V$_G$* plane, for both devices, and observe a single series of well-formed Coulomb diamonds along the *V$_{SD}$* = 0 *V* axis [8]. This observation ensures that the measured current reflects the single-electron tunneling through just one single QD molecule despite the fact that other QD molecules exist in the mesa. The charging energy of the *N*=1 Coulomb diamond is 8 meV (6 meV) for device A (B) (not shown here). These values are smaller than those of single InAs QDs in previous reports [7], because the electronic states in our QD molecules are spread over both dots, and also because the last single QD molecule in the conducting channel is the largest among those inside the mesa.

Figures 1(a) and (b) show gray log-scale *dI/dV$_G$* plots of Coulomb oscillations (*I* vs *V$_G$*) as a function of transverse *B* field whose direction is parallel to current, for devices A and B, respectively. A small *V$_{SD}$* of 0.5 mV is applied to probe just the GSs. We observe spin-paired peaks due to the lifting of the spin degeneracy in each orbital state. In the low *B* field region, several upward kinks reflecting the GS transitions are observed. In order to analyze the evolution of spin-paired peaks, we initially use a standard F-D approach [9], assuming that the dots are laterally confined by a 2D harmonic potential. The F-D state energy is given by $E_{n,l} = (2n + |l| + 1)\hbar\sqrt{\omega_0^2 + \omega_c^2/4} - l/2\,\hbar\omega_c$, where *n* and *l* are the radial and angular momentum quantum numbers, respectively, and $\omega_0$ and $\omega_c$ are the bare confinement potential and cyclotron frequencies, respectively [10]. Figure 1(c) shows the calculated F-D spectrum corresponding to *a single dot* with $\hbar\omega_0$ =10 meV, whose value is estimated for both devices by fitting the *N*=1 Coulomb peak with $E_{0,0}$ [11]. By comparing the data of Fig. 1(a) ((b)) with the calculations, we find that the paired peaks are reproduced by the F-D states from $N=1$ up to 6 (2) for device A (B). This indicates that in Fig. 1(a) ((b)) the S states are consecutively filled by six (two) electrons, starting from the lowest-energy molecular orbital state.

We now compare the $B$ dependencies of the current peaks observed for both devices with those predicted by numerical exact diagonalization based on a single-particle F-D basis set, in order to identify the electronic configurations. In this calculation, we assume that the QD molecule can be treated as two vertically coupled harmonic QDs. The effect of strain is neglected, and only the lowest-energy S and AS states are considered. Details of the calculation are described in Refs. [12] and [13].

Figure 2(a) shows the calculated energy spectra for the $N=1-8$ GSs as a function of transverse $B$ field, corresponding to device A. The GSs (excited states (ESs)) are shown by the solid (dotted) lines. Black and red respectively indicate configurations with no filled AS states or with one or more filled AS states. The boxes and arrows inside represent the S (or AS) states and the spins, respectively. The choice of the S-AS splitting value $\Delta_{SAS} = 10.3$ meV is discussed below. The transition in the $N=4$ GS at $B \cong 0$ T comes from the effect of Hund's first rule, which is however difficult to be distinguished in the experimental data of Fig. 1(a). Several orbital transitions are observed for the $N=5-8$ GSs in the calculation. To compare with the experiment, we show in Fig. 2(b) magnified plot of the Coulomb peaks, which correspond to the $N=5-8$ GSs, as a function of transverse $B$ field, measured for device A. The upward cusps indicated by filled triangles for $N=5-6$ current peaks correspond to the single-particle level crossings between $E_{0,-1}$ and $E_{0,2}$, i.e. between GS configurations $(1sS)^2(2p_+S)^2(2p_-S)$ and $(1sS)^2(2p_+S)^2(3d_+S)$ for $N=5$, and $(1sS)^2(2p_+S)^2(2p_-S)^2$ and $(1sS)^2(2p_+S)^2(3d_+S)^2$ for $N=6$. Here, we represent each orbital state as $(kX)^i$, where $k = 1s$, $2p_+$, $2p_-$, $3d_+$, $4f_+$ stand for $n = 0$ and $l = 0, +1, -1, +2, +3$, respectively, $X=$ S (AS) refers to the S (AS) states, and $i=1$ (2) means single (double) occupancy. This single-particle level crossing corresponds to the beginning of filling factor $v = 2$.

The irregular behavior of the $N$=7-8 peaks in Fig. 2(b) is well explained by the calculation of Fig. 2(a). The weak $B$ dependence of the $N$=7 ($N$=8) GS up to $B \cong 1$ T ($B \cong 0.5$T indicated by the filled triangle) in Fig. 2(b) comes from the filling of the AS $1s$ orbital state, i.e. $(1s\text{S})^2(2p_+\text{S})^2(2p_-\text{S})^2(1s\text{AS})$ $((1s\text{S})^2(2p_+\text{S})^2(2p_-\text{S})^2(1s\text{AS})^2)$. The open triangles show a one-to-one correspondence of the kinks between Figs. 2(a) and (b). The cusps, indicated by blue triangles in Fig. 2(b), occur at lower values of $B$ than those in Fig. 2(a). This implies that the single-particle level spacing between the $3d$ and $4f$ orbital states is smaller than that between the $1s$ and $2p$ or the $2p$ and $3d$ orbital states in the actual device. The filling of the AS $1s$ orbital state does not occur for $N < 7$ because the kinetic energy gain $\hbar\omega_0 - \Delta_{\text{SAS}}$ does not compensate the Coulomb repulsion between S and AS states. The molecular states with filled AS states at $B = 0$ T are present only for $N \geq 7$, when the kinetic energy is strongly lowered by the filling of AS states.

Figure 3(a) shows the calculation for the $N$=1-7 GSs for device B. In this calculation, we use $\Delta_{\text{SAS}} = 3.1$ meV, which is close to the value experimentally estimated from measurements of the $N$=1 excitation spectrum [14]. The $N$=1-2 ($N$=3-4) GSs are the states of $(1s\text{S})$ and $(1s\text{S})^2$ $((1s\text{S})^2(1s\text{AS})$ and $(1s\text{S})^2(1s\text{AS})^2)$. Therefore, molecular states with filled AS states at $B = 0$ T are present for $N \geq 3$. For $N$=5-7 at $B = 0$ T, the S $2p$ orbital states are consecutively filled, which is consistent with the calculation (see Figs. 1(b) and 3(a)). The molecular states with filled AS states for $N \geq 3$ persist to high $B$ field region unlike those in device A because $\Delta_{\text{SAS}}$ in device B is significantly smaller than $\hbar\omega_0$. Figure 3(c) shows a color contour plot of $dI/dV_{SD}$ intensity for the $N$=6-7 peaks measured with $V_{SD} = 4$ mV. We distinguish some GSs and ESs, and their transitions with $B$ field. In the calculated spectrum for $N$=6 we predict transitions indicated by yellow triangles at $B \cong 0$ T and 6.5 T (Fig. 3(a)), but experimentally we can distinguish only the high-$B$ transition (Fig. 3(c)). The triplet-singlet transition at $B \cong 0$ T is related to Hund's first rule, while that at $B \cong 8$ T occurs between two molecular states, i.e. $(1s\text{S})^2(2p_+\text{S})^2(1s\text{AS})^2$ and $(1s\text{S})^2(2p_+\text{S})^2(3d_+\text{S})(1s\text{AS})$. Such molecular state transition can also be characterized by a so-called *isospin* flip, i.e. a change of isospin index $I_s = (N_S - N_{AS})/2$, where $N_S$ ($N_{AS}$) is the total number of electrons occupying S (AS) states [15]. For example, $I_s$ changes from 1 to 2 in the above mentioned transition. For $N=6$, we also identify in Fig. 3(c) a transition between the ESs marked by an open circle at $B \cong 1.5$ T, corresponding to the single-particle level crossing between $E_{0,-1}$ and $E_{0,2}$, i.e. between configurations $(1sS)^2(2p_+S)(2p_-S)(1sAS)^2$ and $(1sS)^2(2p_+S)(1sAS)^2(2p_+AS)$. This transition also occurs in the calculation (see open circle). For $N=7$, we identify the two transitions labeled by yellow triangles. The transitions correspond to changes in molecular GS from $(1sS)^2(2p_+S)^2(2p_-S)(1sAS)^2$ to $(1sS)^2(2p_+S)^2(1sAS)^2(2p_+AS)$ at $B \cong 1.5$ T, and from $(1sS)^2(2p_+S)^2(1sAS)^2(2p_+AS)$ to $(1sS)^2(2p_+S)^2(3d_+S)(1sAS)^2$ at $B \cong 3.5$ T. In these two transitions, first $I_s$ changes from 3/2 to 1/2, and then from 1/2 to 3/2.

Application of an in-plane $B$ field whose direction is perpendicular to current imposes magnetic confinement on electrons in the growth direction, and therefore makes S-AS splitting effectively smaller. This splitting is predicted by perturbation theory to decrease quadratically with in-plane $B$ field [16], whereas Zeeman splitting increases linearly with $B$ field. This different behavior can be used to discriminate between the filling of S and AS states. Figure 3(b) shows the relative spacing $\Delta E_{N+1,N}$ between two neighboring peaks for $N=1$-4 as a function of in-plane $B$ field at 1.5 K, measured for device B. The change of $\Delta E_{2,1}$ is almost proportional to $B$ field, reflecting the effect of Zeeman splitting. We use these data to estimate the electron $g$-factor ~ 0.97 [17]. This value is in good agreement with that obtained by optical spectroscopy [18]. The change of $\Delta E_{3,2}$ and $\Delta E_{5,4}$ is much greater than that of $\Delta E_{2,1}$, and can compare with the $B^2$ fitted lines in red. This observation supports our assignment of consecutive filling of S and AS $1s$ orbital states [19].

While the features observed in device A (B) for $N=6$ ($N=7$) are well reproduced in other devices made from the same material, the behavior for $N \geq 7$ ($N \geq 8$) is not always the same because slight $\Delta_{SAS}$ differences make change significantly the filling sequence. The value of $\Delta_{SAS}$ for device A cannot be directly accessed in the measurement. Nevertheless, the experimental data are in good agreement with the calculation using $\Delta_{SAS}$ = 10.3 meV. Indeed, this value is consistent with recent exact calculations, which take into account realistic dot shape and strain field [20]. The value of $\Delta_{SAS}$ used in the calculation for device B is consistent with its experimental average value estimated from measurements of the *N*=1 excitation spectrum [14] in devices fabricated using material B. This value of $\Delta_{SAS}$ is slightly larger than that predicted in the calculation of Ref. [20].

In the InAs coupled dot system, it is well-known that the size of the upper and lower dots is usually different due to self-assembly process [5]. This effect can give rise to energy offset between the dots [21]. In case that the energy offset is larger than the S-AS splitting, electron filling would be strongly modified due to electron localization in one dot. However, we have not observed any evidence for this.

In conclusion, we have studied the molecular states in single InAs SAD molecules, using single-electron tunneling spectroscopy. We identify the orbital and spin configurations of SAD molecules holding just a few electrons, and confirm different electron filling patterns of the AS states of the QD molecule with weak inter-dot coupling strength. We also observe magnetic field induced transitions between molecular states in the relatively weak magnetic field range.

The authors thank M. Stopa, S. Amaha, T. Hatano, Y. Yamada, Y. Igarashi, K. Ono, and D. G. Austing for valuable discussions. Part of this work is financially supported by the DARPA grant no. DAAD19-01-1-0659 of the QuIST program, SORST-JST, the Grant-in-Aid for Scientific Research A (No. 40302799) and Focused Research and Development Project for the Realization of the World's Most Advanced IT Nation, IT Program, MEXT, MIUR-FIRB RBAU01ZEML, MIUR-COFIN 2003020984, Iniziativa Trasversale INFM Calcolo Parallelo 2005, Italian Ministry for Foreign Affairs – General Bureau of Cultural Promotion and Cooperation (MAE DGPCC).

References

[1] For example, M. Bayer *et al*., *Nature* **405**, 923 (2000).

[2] T. Maltezopoulos *et al*., *Phys. Rev. Lett.* **91**, 196804 (2003).

[3] K. H. Schmidt *et al*., *Phys. Rev. B* **62**, 15879 (2000).

[4] T. Ota *et al*., *Phs. Rev. Lett.* 93, 066801 (2004).

[5] Q. Xie *et al*., *Phys. Rev. Lett*. **75,** 2542 (1995).

[6] M. Bayer *et al*., *Science* **291**, 451 (2001).

[7] B. T. Miller *et al*., *Phys Rev. B* **56,** 6764 (1997) ; M. Fricke *et al.*, *Europhys. Lett*. **36**, 197 (1996).

[8] T. Ota *et al*., *Physica E* **19**, 210 (2003).

[9] V. Fock, *Z. Phys*. **47**, 446 (1928); C.G. Darwin, *Proc. Cambridge Philos. Soc.* **27**, 86 (1930).

[10] The quantum numbers for the F-D states used in this paper are the same as those in the previous experiment given by Ref. 7.

[11] The value of $\hbar\omega_0$ is extracted by fitting $E_{0,0} = \hbar\sqrt{\omega_0^2 + \omega_c^2/4}$ to the $N = 1$ peaks, using $m^* = 0.05\, m_0$ (see Ref. 7), where $m_0$ is the free electron mass.

[12] M. Rontani *et al*., *Phys. Rev. B* **69**, 085327 (2004).

[13] In our calculations we assume as input parameters $\hbar\omega_0 = 10$ meV, effective mass $m^* = 0.05\, m_0$, static dielectric constant $\varepsilon = 15.2$, effective giromagnetic factor $g^* = 2.0$, well width $d = 2$ nm, inter-dot distance between the wetting layers 6.5 (11.5) nm for device A (B). The double quantum well potential profile is chosen to give $\Delta_{SAS} = 10.3$ (3.1) meV for device A (B). We use a basis set made of 17 single-particle orbital states to determine the eigenvalue problems of maximum linear size of the order of $1.25 \cdot 10^5$, after full exploitation of Hamiltonian symmetries by means of a parallel Lanczos code.

[14] T. Ota *et al.*, *Superlattices and Microstructures* **34**, 159 (2003).

[15] J. J. Palacios and P. Hawrylak, *Phys. Rev. B* **51**, 1769 (1995). D. G. Austing *et al., Phys Rev. B*

**70**, 045324 (2004)

[16] Y. Tokura *et al.*, *Physica E* **6**, 676 (2000).

[17] This value is obtained by the linear fitting to the experimental data (see red line in Fig. 3(b)), assuming a constant *g* value. The sign of the *g*-factor is not decided in this experiment, but probably positive following previous works (See Ref. 18).

[18] For example, Y. Toda *et al.*, *Appl. Phys. Lett.* **73**, 517 (1998).

[19] We confirm similar behavior of S-AS splitting and obtain similar values for the *g*-factor for device A under in-plane *B* field.

[20] L. He, G. Bester and A. Zunger, *cond-mat/0503492*.

[21] M. Pi *et al., Phys.Rev Lett.* **87**, 066801 (2001).

[Figure captions]

Figure 1

(a) Gray log-scale plot of current as a function of $V_G$ and transverse $B$ field with $V_{SD}$=0.5 mV for device A. (b) Corresponding for device B. (c) Calculated Fock-Darwin spectrum of *a single dot* assuming $\hbar\omega_0$ =10 meV.

Figure 2

(Color) (a) Numerical calculation for device A with $\Delta_{SAS}$ =10.3 meV. The GSs (ESs) are shown by the solid (dotted) lines. Black and red respectively indicate configurations with no filled AS states or with one or more filled AS states. The boxes and arrows inside represent the S (or AS) states and the spins, respectively. (b) Gray log-scale plot of magnified *N*=5-8 current peaks for device A, derived from Fig. 1(a), with $V_{SD}$ = 0.5 mV as a function of $V_G$ and transverse $B$ field. Filled, open, and blue triangles indicate GS transitions described in the text, kinks, and cusps, respectively.

Figure 3

(Color) (a) Numerical calculation for device B with $\Delta_{SAS}$ =3.1 meV. The GSs (ESs) are shown by the solid (dotted) lines. Black and red respectively indicate configuration with no filled AS states or with one or more filled AS states. The boxes and arrows inside represent the S (or AS) states and the spins, respectively. (b) Relative spacing $\Delta E_{N+1,N}$ between two neighboring peaks for *N*=1-4 as a function of in-plane *B* field whose direction is perpendicular to current, measured for device B at 1.5 K. The red lines are fitted to $\Delta E_{2,1}, \Delta E_{3,2}$ and $\Delta E_{5,4}$. (c) Color contour plot of $dI/dV_{SD}$ intensity for *N*=6-7 current peaks for device B as a function of $V_G$ and transverse $B$ field with $V_{SD}$ = 4 mV. The white dotted lines are a guide for eye. The yellow triangles (open circle) indicate GS (ES) transitions.

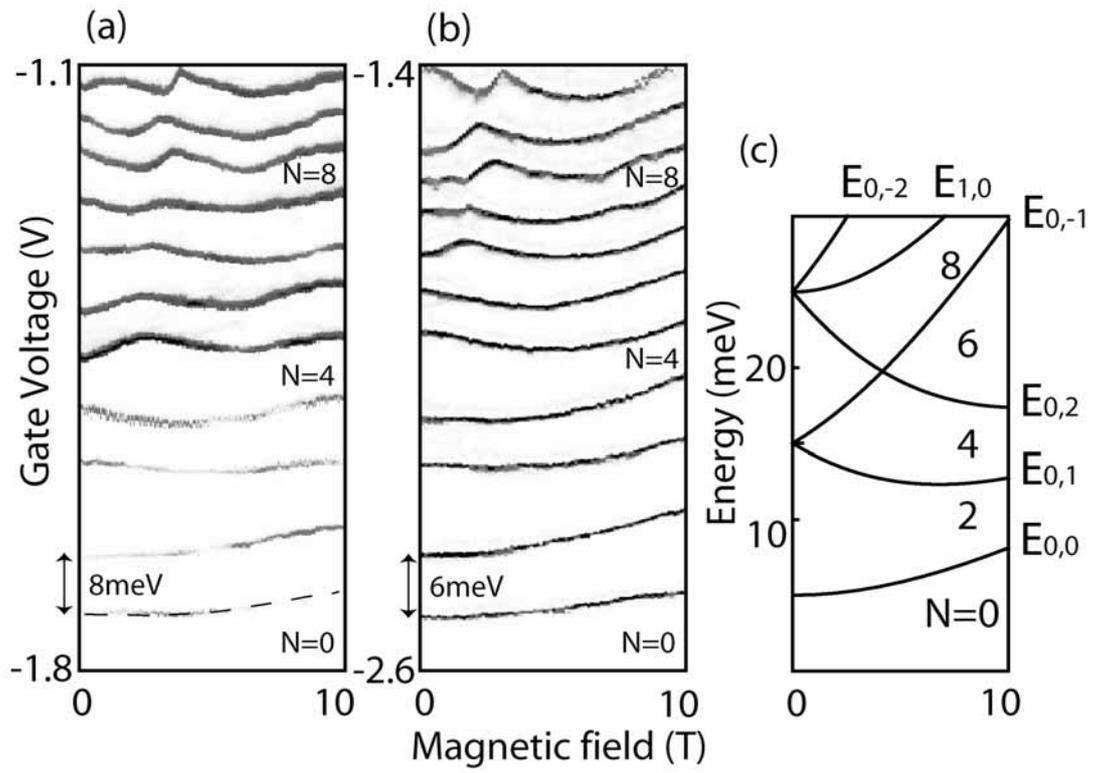

Fig.1

T.Ota et al.

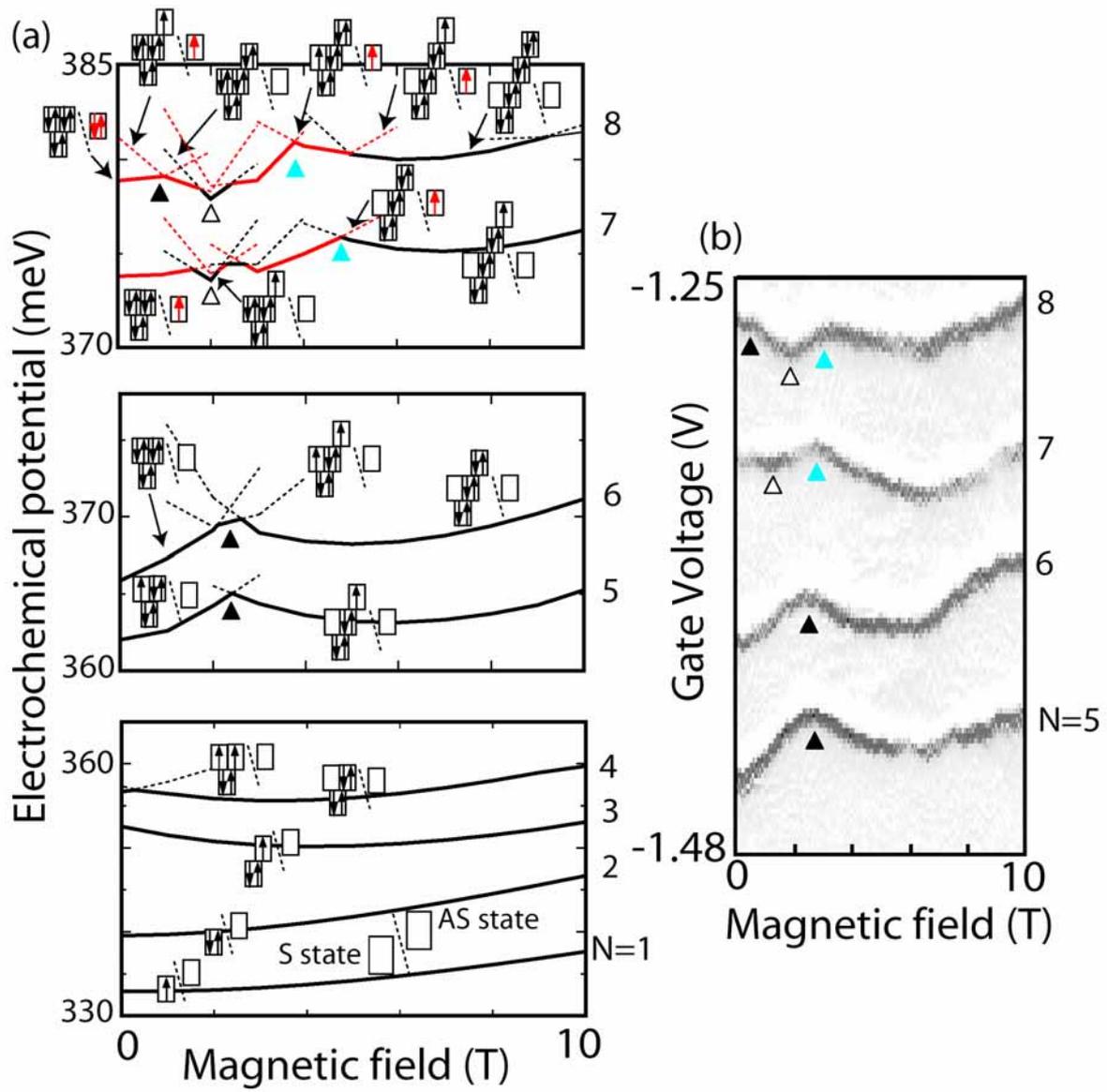

**Fig. 2**

T.Ota et al.

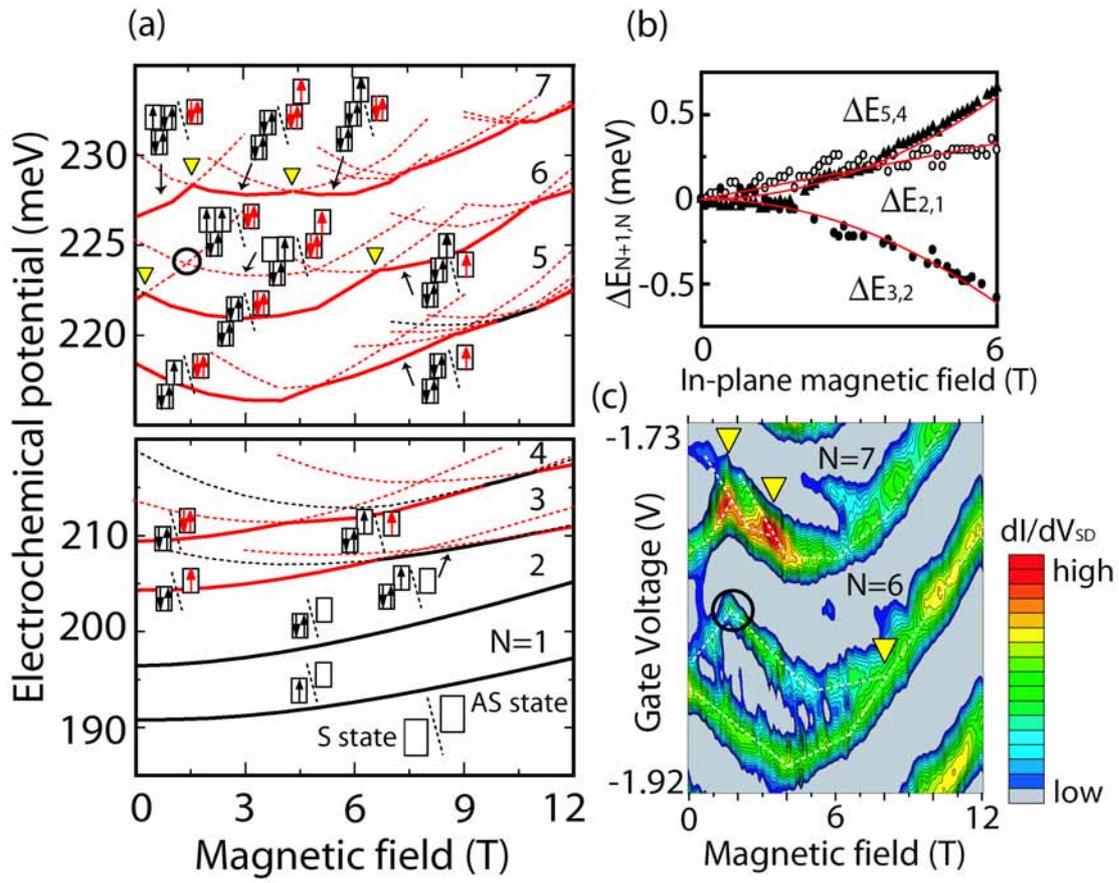

**Fig. 3**

T.Ota et al.